\documentclass[prl,twocolumn,superscriptaddress,nofootinbib,noshowpacs,preprintnumbers,longbibliography,floatfix]{revtex4-2}
\usepackage[utf8]{inputenc}
\usepackage{graphicx}
\usepackage{float}
\usepackage{amssymb}
\usepackage{amsmath}  
\usepackage{mathtools}
\usepackage{dsfont}
\usepackage{array}
\usepackage{bm,fixmath}
\usepackage{mathrsfs}
\usepackage{pifont}
\usepackage{multirow}
\usepackage{upgreek}
\usepackage{xcolor}
\usepackage{bm}
\usepackage{bbm}
\usepackage{physics}
\usepackage{comment}
\usepackage[pdftex,
            pdftitle={Spin Helix},
            pdfauthor={Stefan Kühn, Felix Gerken, Tobias Hartung, Lena Funcke, Paolo Stornati, Karl Jansen, Thore Posske},
            bookmarks,
            colorlinks,
            linkcolor=myblue,
            citecolor=mymagenta,
            menucolor=black,
            urlcolor=myblue,
            plainpages=false,
            pdfpagelabels,
            hypertexnames=false]{hyperref}
\usepackage{verbatim}
\usepackage{slashed}
\usepackage{svg}
\usepackage[resetlabels]{multibib}
\newcites{SM}{SM References}

\definecolor{mymagenta}{RGB}{200, 0, 100}
\definecolor{myblue}{RGB}{45, 48, 146}
\definecolor{mygreen}{RGB}{0, 126, 0}

\newcommand{\thore}[1]{{\color{mygreen}#1}}

\makeatletter
\def\maketitle{
\@author@finish
\title@column\titleblock@produce
\suppressfloats[t]}
\makeatother

\newcommand{\onecolumngridWithRule}{\onecolumngrid
\noindent\rule[0ex]{\linewidth}{1pt}}
\newcommand{\twocolumngridWithRule}{
\noindent\rule[0ex]{\linewidth}{1pt} \twocolumngrid}

\newcommand{\hmax}{\ensuremath{h_\mathrm{max}}}
\renewcommand{\vec}[1]{\bm{#1}}

\newcommand{\refFig}[2][]{Fig.~\ref{#2}#1}

\newcommand{\refEq}[1]{Eq.~(\ref{#1})}

\newcommand{\citeRef}[1]{Ref.~\cite{#1}}

\newcommand{\ourtitle}{Quantum spin helices more stable than the ground state: onset of helical protection}

\renewcommand{\section}[1]{\textit{#1.---}}

\begin{document}
\title{\ourtitle{}}

\author{Stefan K{\"u}hn}
\affiliation{Computation-Based Science and Technology Research Center, The Cyprus Institute, 
Nicosia, Cyprus}
\affiliation{Deutsches Elektronen-Synchrotron DESY,
Zeuthen, Germany}

\author{Felix Gerken}
\affiliation{I. Institut für Theoretische Physik, Universität Hamburg,
Germany}
\affiliation{The Hamburg Centre for Ultrafast Imaging, 
Hamburg, Germany}

\author{Lena~Funcke}
\affiliation{Transdisciplinary Research Area ``Building Blocks of Matter and Fundamental Interactions'' (TRA Matter) and Helmholtz Institute for Radiation and Nuclear Physics (HISKP), University of Bonn, Bonn, Germany}
\affiliation{Center for Theoretical Physics, Co-Design Center for Quantum Advantage, and NSF AI Institute for Artificial Intelligence and Fundamental Interactions, Massachusetts Institute of Technology, 
Cambridge, MA, USA}

\author{Tobias Hartung}
\affiliation{Northeastern University - London,
London, 
UK}

\author{Paolo Stornati}
\affiliation{ICFO-Institut de Ciencies Fotoniques, The Barcelona Institute of Science and Technology, 
Castelldefels (Barcelona), Spain}

\author{Karl Jansen}
\affiliation{Deutsches Elektronen-Synchrotron DESY,
Zeuthen, Germany}

\author{Thore Posske}
\affiliation{I. Institut für Theoretische Physik, Universität Hamburg,
Germany}
\affiliation{The Hamburg Centre for Ultrafast Imaging, 
Hamburg, Germany}

\date{\today}

\preprint{MIT-CTP/5478}

\begin{abstract}
Topological magnetic structures are promising candidates for resilient information storage.
An elementary example are spin helices in one-dimensional easy-plane quantum magnets.
To quantify their stability, we numerically implement the stochastic Schr{\"o}dinger equation and time-dependent perturbation theory for spin chains with fluctuating local magnetic fields.
We find two classes of quantum spin helices that can reach and even exceed ground-state stability: Spin-current-maximizing helices and, for fine-tuned boundary conditions, the recently discovered ``phantom helices''.
Beyond that, we show that the helicity itself (left- or right-rotating) is even more stable. 
We explain these findings by separated helical sectors and connect them to topological sectors in continuous spin systems.
The resulting helical protection mechanism is a promising phenomenon towards stabilizing helical quantum structures, e.g., in ultracold atoms and solid state systems. 
We also identify an---up to our knowledge---previously unknown new type of phantom helices.
\end{abstract}

\maketitle

\section{Introduction}
\label{sec:introduction}
Quantum states are notoriously vulnerable to external perturbations.
Yet, aside from cooling or physically separating quantum systems from the environment, some mechanisms create
comparably stable quantum phenomena. Among these are topological electronic
phases~\cite{Kitaev2009,Schnyder2009change,Hasan2010} including the Quantum Hall effects~\cite{Klitzing1980,Tsui1982,Laughlin1983AnomalousQuantumHallEffectAnIncompressibleQuantumFluidWithFractionallyChargedExcitations,Klitzing86,BernevigHughesZhang2006,Konig2007},
topological superconductors~\cite{Kitaev2001UnpairedMajoranaFermionsInQuantumWires,Ivanov2001NonAbelianStatisticsOfHalfQuantumVortices,Kim2018TowardMajorana,Schneider2021,Schneider2022}, topological spin models~\cite{Kitaev2006AnyonsInAnExactlySolvableModelAndBeyond},
and spin-based anyons~\cite{Kitaev2006AnyonsInAnExactlySolvableModelAndBeyond}. 
Furthermore, quantum systems affected by specific external perturbations can reach dark states, i.e., 
subspaces protected against decoherence~\cite{Plenio1999, Gau2020}.

Recently, helices in easy-plane one-dimensional Heisenberg magnets were conjectured to extend the class of stable quantum states, 
having been predicted to exhibit stability in classical systems~\cite{VedmedenkoAltwein2014TopologicallyProtectedMagneticHelixForAllSpinBasedApplications},
in semi-classical approximations~\cite{KimTakeiTserkovnyak2016}, 
and in quantum systems~\cite{Popkov2013PhysRevE,Popkov2016PhysRevA,Popkov2020,Posske2019}, 
including dissipatively and parametrically controlled magnetic boundaries that facilitate their creation~\cite{PopkovPRL2013ExactMatrixProductSolutionOfBoundaryDissipativeSpinHelices,Popkov2017PhysRevA,Popkov2017JPhysA,PopkovSchuetz2017SolutionOfTheLindbladEquationForSpinHelixStates,Posske2019,Popkov2022}. 
In particular, helical solutions for quantum spin chains with fine-tuned magnetic boundary fields were found, which are product states of spins at individual sites.
Using the Bethe ansatz, these helices were shown to consist of ``Bethe phantom roots''~\cite{Popkov2021},
 which carry zero energy but a finite momentum relative to a reference state~\cite{Nepomechie2003,Nepomechie2003b,Nepomechie2004Addendum,Cao2003,Cao2013,Zhang2021a,Popkov2021}.
Jepsen et.\ al~\cite{Jepsen2022} have demonstrated the creation of such phantom helices in cold atomic systems
and put phantom helices in relation to quantum scars, i.e., states that equilibrate significantly slower than an average state ~\cite{Moudgalya2022}.

Topological spin systems could be used to store energy like in a spring~\cite{VedmedenkoAltwein2014TopologicallyProtectedMagneticHelixForAllSpinBasedApplications}, both in classical and in quantum spintronics, and
as bits and qubits, by storing information in its rotational sense.
To this end, proposals using  quantum skyrmions~\cite{Psaroudaki2021} and quantum merons~\cite{Xia2022} have been made.
Quantum spin helices and quantum spin systems are an active research area in solid states physics~\cite{Kim2018TowardMajorana,ChoiReview2019,Liebhaber2021b} and quantum chemistry~\cite{ZhaoJiangEtAl2022}, and, beyond their realization in ultracold atom systems, could be simulated with tensor networks or on a quantum computer. 
Furthermore, by a Jordan-Wigner transformation, quantum spin helices are closely connected to Josephson junctions which exhibit a helically twisted superconducting order parameter~\cite{Shen2019}, and understanding spin helices may help to analyze higher-dimensional noncollinear quantum magnetism like quantum skyrmions~\cite{Siegl2022} and generalized phantom states~\cite{Jepsen2022}.
In all these contexts, it is paramount to quantitatively understand the susceptibility of quantum spin helices to external noise in the bulk of the chain. Particularly relevant are parametric perturbations, which correspond to, e.g., fluctuating magnetic fields, fluctuating superconducting order parameters, phonons, or gate errors, depending on the physical system at hand.

In this manuscript, we show that quantum spin helices in one-dimensional easy-plane Heisenberg magnets generally exhibit noise protection that can exceed ground state stability. Furthermore, the helicity of a state is protected for even larger time scales.
To show this, we analyze quantum spin chains with random, time-fluctuating on-site magnetic
fields by simulating the stochastic Schr{\"o}dinger equation and 
by employing time-dependent perturbation theory.
Our study includes phantom helices and the more general class of quantum spin helices characterized as the helices carrying maximal spin current along the chain. 
The stability of quantum spin helices is explained by the length-dependent onset of decoupled helical sectors,
 distinguishing \mbox{left-,} \mbox{right-,} and non-rotating quantum spin states.
 We speculate that this helical protection of ferromagnetic quantum spin helices is descending from the topological protection~\cite{SupplementalMaterial} in continuous antiferromagnetic spin systems with large spin quantum numbers~\cite{KimTserkovnyak2016TopologicalEffectsOnQuantumPhaseSlipsInSuperfluidSpinTransport}. 
The helical protection preserves the helicity of a quantum state for short and intermediate time scales and could be a base for future stable helical quantum effects in ultracold atoms and solid state systems. 

\begin{figure}
    \centering
    \includegraphics[width = 1  \linewidth]{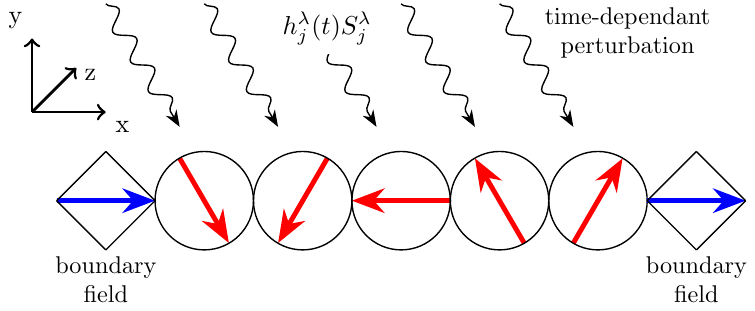}
\caption{\label{fig:system}We consider quantum helices defined by helical spin expectation values
    of a chain of coupled spins or pseudo-spins (red) exposed to uncorrelated time-dependent perturbations of the magnetic field $|{h}^\lambda_{j}(t)|\leq h_\text{max}$.
    The helix is stabilized by boundary fields in $x$-direction (blue).}
    \label{figQubitHelix}
\end{figure}

\begin{figure*}
    \includegraphics[width = 1.0\textwidth]{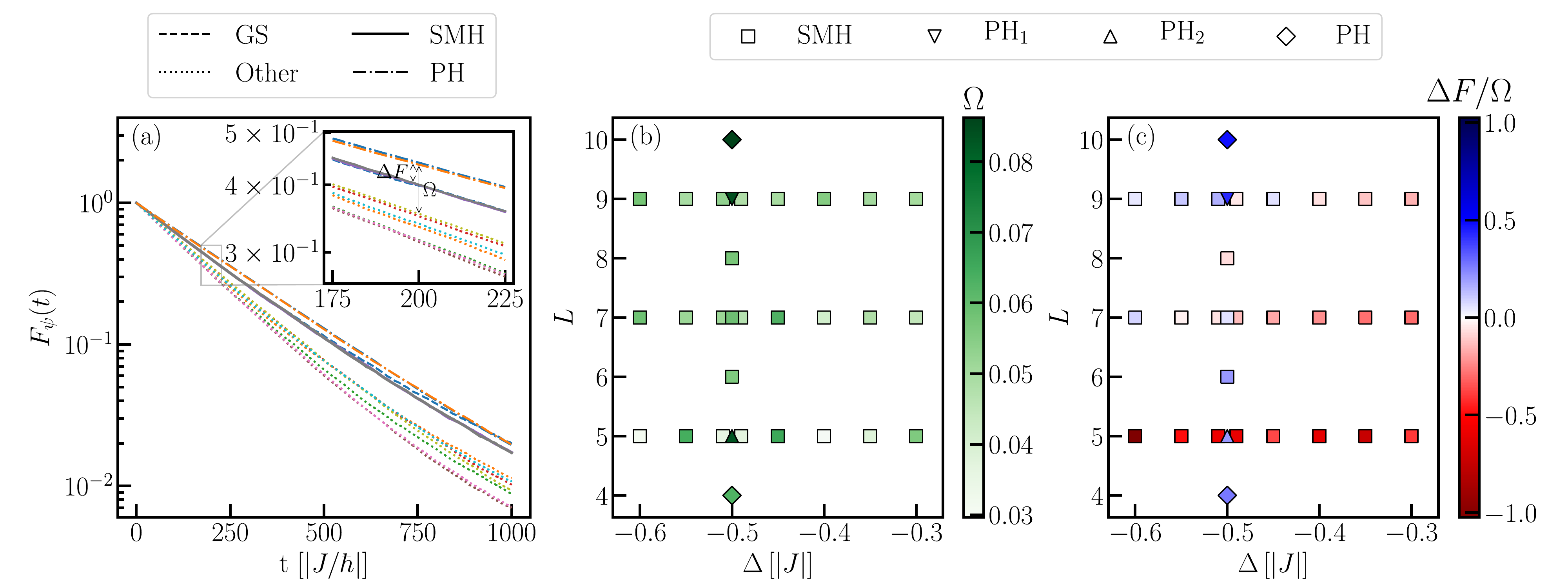}
    \caption{The stability of quantum spin helices. (a)~The ground state (blue dashed line), the two spin-current-maximizing helices (solid lines, degenerate despite numerical fluctuations), and the phantom helices (PH, dash-dotted lines) are significantly more stable than other states, shown by exact simulations of the fidelity $F_{\Psi}(t)$, see Eq.~\eqref{eqn:TimeEvolutionOverlap} and main text. Inset: At time $t_s = 200 \hbar/|J|$, the most stable helical state is separated from the first excited nonhelical state (dashed orange line) by the fidelity difference $\Omega$ and separated from the ground state by the fidelity difference $\Delta F$. 
    Parameters: chain length ${L}=10$, anisotropy $\Delta = J/2$, perturbation strength $\hmax = |J|/2$, averaged over $1000$ runs with different random noise.
     (b) The separation $\Omega$ between the most stable helical state and less stable states at time $t_\text{s}= 200 \hbar/|J|$ (see inset of~\refFig[a]{fig:Stability}) shows that the  most stable helix can be a spin-current-maximizing helix (SMH, square) or a phantom helix of type $1$ with $M=3$ ($\text{PH}_1$, downwards triangle), type $2$ with $M=-1$ ($\text{PH}_2$, upwards triangle), or type $1$ and $2$ with $M=1$ (PH, diamond), respectively.
     (c) Fidelity difference $\Delta F$ between the ground state and the most stable helical state in units of $\Omega$.
     For spin-current-maximizing helices, stability increases in chain length $L$, where chains with $L\gtrapprox 6$ sites can become more stable than the ground state (blue), while phantom helices are more stable than the ground state for all $L$.
     }
     \label{fig:Stability}
\end{figure*}

\section{Noise model for Heisenberg chains}
\label{sec:model}
We consider a one-dimensional easy-plane $XXZ$ Heisenberg magnet of spins $1/2$, which is exposed to time-dependent random fluctuations of local magnetic fields.
The Hamiltonian is 
\begin{align}
\label{eqn:Hamiltonian}
    H(t) &= H_\text{chain} + H_\text{end} + H_{\text{rand}}(t) ,
    \nonumber \\
    H_\text{chain} &= 
        \sum_{j<{L}} 
         J \left(
                S^x_j S^x_{j+1} + S^y_j S^y_{j+1}
            \right)
            + 
              \Delta
            S^z_j S^z_{j+1},
    \nonumber \\
    H_\text{end} &=  J
        \left(
		S_1^x + S_{L}^x
	 \right) \hbar/2,
    \nonumber \\
    H_\text{rand}(t) &= 
        \sum_{{j}\leq {L}, \lambda} 
         h^\lambda_{j}(t) S^\lambda_{j}. 
\end{align}
Here, ${L}$ is the length of the spin chain, $S^\lambda_j$ is the spin operator on the $j^\text{th}$ site in direction $\lambda \in \{x,y,z\}$,
 $J$ is the easy-plane coupling, and $\Delta$ is the axial anisotropy in the $z$-direction. 
 We consider ferromagnetic coupling $J<0$, yet our results directly transfer to planar antiferromagnetism by the mapping ${S_j}^x, {S_j}^y \to  - {S_j}^x, -{S_j}^y$ for even $j$.
The magnitude of the boundary fields in $H_\text{end}$~\cite{Posske2019,Zhang2021a} is generic in the sense that a large magnetic field at hypothetical sites $0$ and ${L}+1$, which fully polarize these spins, will create exactly the desired magnitude of the boundary fields. 
The coupling constant $h_{j}^\lambda(t)$ fluctuates randomly in time between between $\pm h_{\text{max}}$ and is uncorrelated for different lattice sites. 
For simplicity, we assume that the perturbations change stroboscopically in intervals of $\delta t$.
This approach is also a first step towards simulating~\refEq{eqn:Hamiltonian} on a noisy quantum computer, where Trotter real-time evolution along with a decomposition of the single-step time evolution operator in quantum gates could be employed. In this scenario, uncorrelated coherent single-qubit gate errors correspond to the random parametric noise assumed here.
Within the Markovian approximation, neither the assumed uniform distribution nor the sudden changes of the magnetic field cause unphysical behavior in the limit of small $\delta t \dot J/\hbar$. This is demonstrated by the corresponding Lindblad master equation that assumes the form of a usual continuous Markovian time evolution of the system with uncorrelated external fields~\cite{SupplementalMaterial}. The Lindblad superoperator does not assume a known integrable form~\cite{DeLeeuwEtAl2021ConstructingIntegrableLindbladSuperoperators}, including the coupling of the fluctuating magnetic fields to noncommuting operators.

We call a quantum state a quantum spin helix if the expectation values of the local spins form a helix in the \mbox{$x-y$} plane, see~\refFig{fig:system}.
In general, a helical eigenstate of $H_\text{chain}$ is degenerate to a state with opposite helicity 
 because $H_\text{chain}$ has the symmetry $U=\prod_{j\leq {L}} S^x_j$ that mirrors each spin about the $x$-axis.
Quantum spin helices are therefore ambiguously defined when only considering their energy.
To resolve the ambiguity, we consider two special kinds of helices. 
First, helices that are eigenstates of the model in \refEq{eqn:Hamiltonian} with the maximal amount of spin current along the direction of the chain $C = \sum_{j=1}^{L} \left(\vec{S}_j \cross \vec{S}_{j+1}\right)^z$.
These spin-current-maximizing helices consist of entangled spins and generally appear in chains of odd length, or at special anisotropies $\Delta = J\cos(\pi/{k})$ for odd ${k}<{L}$ with even chain lengths $L$~\cite{Posske2019}. Here, we focus on $\Delta = J/2$, i.e., ${k}=3$. Spin-current-maximizing helices at general easy-plane values of $\Delta$ $\left(|\Delta|<|J|\right)$ can be prepared by adiabatically twisting the boundary magnetization by $2\pi$ for $\Delta=J/2$~\cite{Posske2019} and subsequently adiabatically adjusting to the desired value of $\Delta$.
The second class of quantum spin helices appears when the chain length ${L}$ and the Heisenberg anisotropy $\Delta$ matches the phantom condition
 \mbox{$\left({L} - M\right)\gamma  +\delta_{M,1} \pi \equiv 0 \ (\operatorname{mod} 2\pi)$}
 with \mbox{$\gamma = \arccos{\Delta/J}$} and $M$ being $-1$, $1$, or $3$.
Then, helices with constant winding angle $\gamma$ are product states of local spin states fulfilling the phantom helix ansatz~\cite{Cao2003,CerezoRossignoliCanosa2016,Zhang2021a,Popkov2021},
\begin{align}
\label{eqn:Phantom}
 \ket{\mathrm{PH}_\tau} = \bigotimes_{j=1}^{L} R_{z}
    		\left(
        		\pm
			 \left[
            			j - 2 \delta_{\tau,1}
            		\right]
        		\gamma + \pi \delta_{\tau,1}
    		\right)
	\ket{\rightarrow}_j.
\end{align}
Here, the index $\tau = 1, 2$ denotes two types of phantom helices, $R_z\left(\theta\right)$ is an $SU(2)$ rotation around the $z$-axis 
 with the angle $\theta$, and $\ket{\rightarrow}_j$ is the spin state at site $j$ pointing into the $x$-direction.
Type $\tau=1$ helices are eigenstates for $M=1,3$ and type $\tau=2$ helices are eigenstates for $M=1,-1$. 
The cases where $M=1$ is---up to our knowledge---a previously unknown phantom condition~\cite{Cao2003,Cao2013,Zhang2021a,Popkov2021}, fulfilling the criteria of~\citeRef{CerezoRossignoliCanosa2016},
which we verified by acting with $H_{\text{chain}}$ on the ansatz above.
To prepare phantom helices, an initial product state gets twisted locally by single-spin manipulations~\cite{Jepsen2022}.

\section{Stability of quantum spin helices\label{sec:results}}
In the following, we present results on the stability of the helical and nonhelical eigenstates of $H_\text{chain}$ for varying chain length and Heisenberg anisotropy $\Delta$. 
For the boundary fields in~\refEq{eqn:Hamiltonian}, spin-current-maximizing helices exist for even chain lengths with $\Delta = J/2$ and for odd chain lengths for all $|\Delta|<|J|$. For the considered value of $\Delta=J/2$, phantom helices exist only for chain lengths ${L}=3,4,5,9,10,11, \dots$, 
 i.e., ${L} \equiv 3 ,4,$~or~$ 5\  (\text{mod } 6)$, inferred from the phantom condition and \refEq{eqn:Hamiltonian}.
 
First, we numerically simulate the Hamiltonian, implementing the time evolution using a series expansion for the time evolution operator up to second order in the time step $\delta t$, which we choose to be $0.01 \hbar/J$, and average over $1000$ runs.
Other values of $\delta t$ or higher-order terms in the expansion do not change our results qualitatively.
\begin{figure*}\includegraphics[width=1.0\textwidth]{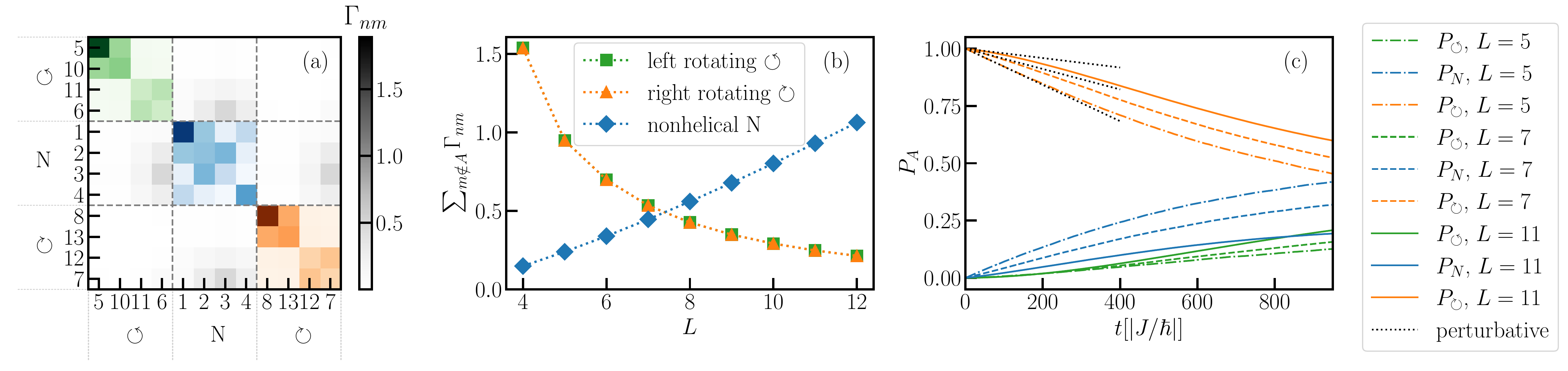}
    \caption{Onset of helical protection with increasing chain length.
    (a)	Formation of helical sectors: reduced transition amplitudes $\Gamma_{\psi, \psi^\prime}$ appear between left- ($\circlearrowleft$, green), right- ($\circlearrowright$, red),
	and nonhelical ({N}, blue) eigenstates of $H_{\text{chain}}(t=0)$. Only the four energetically lowest states are shown for each sector, sorted according to increasing absolute value of the spin current. Data corresponds to ${L}=10$ and $h_\text{max} = |J|/2$.
    (b) Decreasing transition amplitudes away from the helical sectors  with increasing chain lengths $L$ for the energetically lowest state in each sector (triangle and square markers). In contrast, the nonhelical ground state (diamond) experiences increased transitions with increasing $L$. Shown is the sum of the matrix elements $\sum_{m\notin A}\Gamma_{nm}$.
    (c)~The length-dependent preservation of helical sectors spanned by left-rotating helical states ($\circlearrowleft$, green), right-rotating helical states ($\circlearrowright$, orange) and nonhelical states ($N$, blue) for the system sizes ${L}=7$ (dash-dotted lines), ${L}=9$ (dashed lines) and ${L}=11$ (solid lines). Dashed gray lines denote the perturbative result, see \refEq{eqn:perturbationTheorySectors}. $P_A$ is the probability of finding the time-evolved initial state, a right-rotating spin-current-maximizing helix, after time $t$ in the left- ($\circlearrowleft$, green), right- ($\circlearrowright$, orange), or nonhelical sector ($N$, blue). Dashed lines denote the pertubative results for $P_A$, see \refEq{eqn:perturbationTheorySectors}. Parameters: chain length ${L}=10$, anisotropy $\Delta = J/2$, perturbation strength $\hmax = |J|/2$, averaged over $1000$ runs with different random noise.
    }\label{fig:time_scale_separation}
\end{figure*}
As a measure for stability, we consider the expected fidelity of an eigenstate of the fluctuation-free Hamiltonian and a time-evolved initial eigenstate 
\begin{align}
\label{eqn:TimeEvolutionOverlap}
    F_{\Psi}(t)=
        \mathbb{E}\left(|\langle \Psi(0)| \Psi(t) \rangle|^2\right),
\end{align}
where $\mathbb{E}\left(\dots \right)$ denotes averaging with respect to the random noise.

We find that the ground state, the spin-current-maximizing helices, and the phantom helices, in case they exist,
 are more stable than the remaining eigenstates, see~\refFig[a]{fig:Stability}.
A suitable time for comparing chains of different length is $t_\text{s} = 200  \hbar/|J|$,
 where the fidelities usually reach $F_\Psi(t_\text{s}) \approx 40\%$ and the group of stable states is separated from less stable ones, see \refFig[a]{fig:Stability} for a representative example with chain length $L=10$.
 To elaborate this, we consider the fidelity difference $\Omega = F_{\Psi_\text{h}}(t_s)-F_{\Psi_\text{nh}}(t_s)$ between the most stable helical state $\Psi_\text{h}$ and the most stable nonhelical excited state $\Psi_\text{nh}$. As a general trend, $\Omega$ increases for increasing chain length, and the most stable helical state is consistently separated from the most stable nonhelical excited state, independent from the Heisenberg anisotropy, see \refFig[b]{fig:Stability}.

 To show that helical states can exceed ground state stability, we calculate the difference in fidelity between $\Psi_\text{h}$ and the ground state $\Psi_{\text{gs}}$, $\Delta F = F_{\Psi_\text{h}}(t_s)-F_{\Psi_\text{gs}}(t_s)$. The sign of $\Delta F$ determines parameter regions where either the ground state or $\Psi_\text{h}$ is most stable, see the red and blue regions in ~\refFig[c]{fig:Stability}, respectively. 
We find that in case phantoms exist, they are the most stable states, independently of the length $L$ of the chain and the satisfied phantom condition.
For spin-current-maximizing helices, the stability increases with $L$, with a turning point around $L\approx 6$, where spin-current-maximizing helices can become more stable than the ground state in dependence on the Heisenberg anisotropy.

\section{Onset of helical sectors}
Despite their enhanced stability,
we observe from~\refFig{fig:Stability}
that quantum spin helices are evidently not perfectly stable.
We find that albeit the initial states experience strongly reduced transitions to states with different helicity, transitions to states with the same helicity are not suppressed.
This decoupling of the helical sectors becomes more prominent for longer chains.
To elaborate, we consider the probability $P_{A}(n,{t}) = \sum_{g \in A} |\langle g|n({t})\rangle|^2$ of measuring the system at time $t$ in the same helical sector \mbox{$A\in \{\circlearrowleft \text{left-rotating},\ \circlearrowright \text{right-rotating},\ $N$ \text{ (nonhelical)}\}$} as the initial state $|n\rangle$. Here, left-rotating, right-rotating, and nonhelical states are defined by positive, negative, and vanishing spin current, respectively. 
For $h_\text{max} \cdot \delta t / \hbar \ll 1$ and $\delta t \cdot J/\hbar \ll 1 $, we find~\cite{SupplementalMaterial}
\begin{align}
P_{A}(n,{t}) \approx
\begin{cases}
 1-\frac{ \delta t\cdot h_\text{max}^2}{3\hbar^2} {t} \sum\limits_{m \not \in A} 
 \Gamma_{nm} & \text{$n \in A$} ,
 \\
 \frac{\delta t \cdot h_\text{max}^2  }{3\hbar^2} {t} \sum\limits_{m \in A} 
 \Gamma_{nm} & \text{$n \not \in A$,} 
\end{cases}
\label{eqn:perturbationTheorySectors}
\end{align}
with $n$ and $m$ labeling the eigenstates of \mbox{$H_\text{chain}({t}=0)$}. 
The matrix $\Gamma_{nm} = \sum_{j=1}^{{L}} \sum_{\lambda} \abs{\bra{m} S^\lambda_j\ket{n}}^2$ is the relevant quantity for describing transitions
 between states of different helical sectors and shows a strong separation between helical sectors, as depicted in~\refFig[a]{fig:time_scale_separation}.
The transition from one helical sector to another one is proportional to $\Gamma_A(n) = \sum_{m \notin A} \Gamma_{nm}$. 
In~\refFig[b]{fig:time_scale_separation}, we show the dependence of $\Gamma_\circlearrowleft$, $\Gamma_\circlearrowright$, and $\Gamma_N$ for the energetically lowest spin-current-maximizing helices and the ground state on the chain length $L$. The short-time decay of the spin-current-maximizing helices falls below the one of the ground state at intermediate lengths, and ultimately approaches zero. This is remarkable, because longer chains contain more disorder terms, such that a faster decay of general properties of the states is expected, as observed in the increasing short-term decay of the ground state $\Gamma_N$. Notice that the helicity of the ground state during the stochastic time evolution remains zero by decaying with equal probability into the right- and left-rotating sectors.
This implies that, for sufficiently short times  ${t}$ and long chains, the spin-current-maximizing helices only decay into states in the same helical sector.
For longer times, 
leaving the perturbative regime, this tendency prevails within an intermediate time regime whose width depends on the strength of the random noise $h_{\text{max}}$, as shown by the increase of  $P_\circlearrowleft(t)$ for chains lengths $7$, $9$, and $11$ in~\refFig[c]{fig:time_scale_separation}, representatively for $h_{\text{max}} = \Delta = J/2$, using $1000$ independent runs of the full time evolution. Until ${t}\approx 200 \hbar/|J|$, the increase in chain length causes a net protective effect against transitions to the oppositely rotating and to the nonhelical sector, as seen by the decreased absolute slope of $P_\circlearrowright$ for small times. For longer times, the slopes of $P_\circlearrowright$ for different chain lengths become similar, and the population of the  oppositely rotating helical sector (green) is no longer negligible, which implies that helical protection is lost.

The decoupling between sectors of different helicities is reminiscent of topological sectors in continuum theories for large spin quantum numbers. There, 
a semiclassical saddle point analysis for antiferromagnetic helices~\cite{KimTserkovnyak2016TopologicalEffectsOnQuantumPhaseSlipsInSuperfluidSpinTransport}
revealed that spin slips, the only causes of transitions between helical sectors, are strongly suppressed by a topological $\theta$-term
in the action. Interestingly, we find that the stability of helices is the same for ferromagnetic models, where, instead, the suppression of spin slips is caused by a topological Wess-Zumino-Witten term~\cite{altland_simons_2010,SupplementalMaterial}.
Due to the Wess-Zumino-Witten term, spin flips with opposite skyrmion charge destructively interfere in the case of half-odd integer spin systems,
as discussed in~\cite{SupplementalMaterial}.

\section{Discussion}
\label{sec:discussion}
This study reveals the onset of decoupled helical sectors in spin chains of finite length when the chain is perturbed by local randomly fluctuating magnetic fields.
The resulting helical protection increases for increasing chain lengths and suppresses transitions to sectors of different helicities for short and intermediate time scales.
We suggest that helical protection becomes weaker at longer time scales due to stronger coupling between high-excited states, which facilitates transitions to sectors with a different helicity. In case these transitions were suppressed by additional measures, e.g., by a low-temperature bath, we expect the helical sectors to display their stability over an increased time-scale. In general, such a time-scale separation and decoupled states would be a hallmark feature of weak ergodicity breaking~\cite{Moudgalya2022}. 

While the transitions between the helical sectors are strongly suppressed, states experience no native protection against excitations within a helical sector.
This underlines the challenge in using quantum spin helices, quantum skyrmions~\cite{Psaroudaki2021}, or quantum merons~\cite{Xia2022} as qubits, where a combination with
conventional quantum error correction or mitigation techniques would need to be applied to suppress these unwanted transitions. In future research, we aim to investigate whether the helical protection mechanism itself could be useful for quantum computing applications.

\section{Acknowledgments}
The authors thank Se Kwon Kim for discussions and  Balz{\'a}s Pozsgay for remarks, TP and FG thank Rafael Nepomechie for discussions, and TP thanks Mircea Trif 
for discussions.
SK acknowledges financial support from the Cyprus Research and Innovation Foundation under projects
 ``Future-proofing Scientific Applications for the Supercomputers of Tomorrow (FAST)'', contract no.\ COMPLEMENTARY/0916/0048,
 and ``Quantum Computing for Lattice Gauge Theories (QC4LGT)'', contract no.\ EXCELLENCE/0421/0019.
FG acknowledges funding by the Cluster of Excellence ``CUI: Advanced Imaging of Matter'' of the Deutsche Forschungsgemeinschaft
 (DFG) - EXC 2056 - project ID 390715994. 
LF is partially supported by the U.S.\ Department of Energy, Office of Science, National Quantum Information Science Research Centers,
Co-design Center for Quantum Advantage (C$^2$QA) under contract number DE-SC0012704, by the DOE QuantiSED Consortium under subcontract number 675352,
by the National Science Foundation under Cooperative Agreement PHY-2019786
 (The NSF AI Institute for Artificial Intelligence and Fundamental Interactions, \url{http://iaifi.org/}),
and by the U.S.\ Department of Energy, Office of Science, Office of Nuclear Physics under grant contract numbers DE-SC0011090 and DE-SC0021006. PS acknowledges support from Ministerio de Ciencia y Innovation Agencia Estatal de Investigaciones (R\&D project CEX2019-000910-S, AEI/10.13039/501100011033, Plan National FIDEUA PID2019-106901GB-I00, FPI), Fundació Privada Cellex, Fundació Mir-Puig, and from Generalitat de Catalunya (AGAUR Grant No. 2017 SGR 1341, CERCA program), and MICIIN with funding from European Union NextGenerationEU(PRTR-C17.I1) and by Generalitat de Catalunya. TP acknowledges funding by the DFG (project no.\ 420120155). 
This work is funded by the European Union's Horizon Europe Framework Programme (HORIZON) under the ERA Chair scheme with grant agreement No. 101087126.
This work is supported with funds from the Ministry of Science, Research and Culture of the State of Brandenburg within the Centre for Quantum Technologies and Applications (CQTA).
\begin{center}
    \includegraphics[width = 0.05\textwidth]{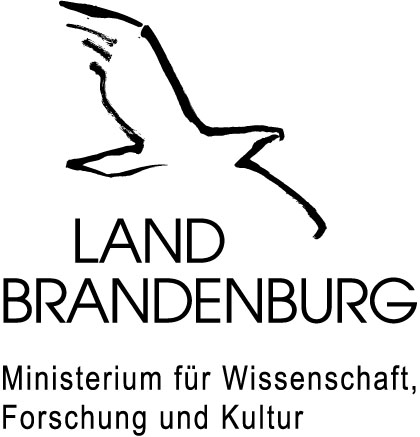}
\end{center}

\bibliographystyle{apsrev4-1}
\bibliography{Papers,library}


\clearpage


\setcounter{equation}{0}
\setcounter{figure}{0}
\setcounter{table}{0}

\makeatletter
\renewcommand{\theequation}{S\arabic{equation}}
\renewcommand{\thefigure}{S\arabic{figure}}
\renewcommand{\bibnumfmt}[1]{[S#1]}
\renewcommand{\citenumfont}[1]{S#1}
\title{Supplemental Material: \ourtitle{}}
\maketitle

\section{Topological stability of ferromagnetic quantum spin helices: nonlinear sigma model}
For antiferromagnetic spin helices, Ref.~\citeSM{Kim_2016} demonstrated that quantum phase slips (QPS) unwind spin helices with a total winding angle of $\Delta\phi = 2\pi$ for integer spins, assuming the large-spin and continuum limits. For half-odd integer spins, these QPS destructively interfere, such that the helices remain stable. In this appendix, we show that the same stability arguments hold true for ferromagnetic spin helices.

We start with a brief summary of the results obtained in Ref.~\citeSM{Kim_2016}, which consider the Hamiltonian of the anisotropic Heisenberg antiferromagnetic spin-$s$ chain
\begin{equation}
    H = J \sum_n \left[ \mathbf{S}_n \cdot \mathbf{S}_{n+1} - a S_n^z S_{n+1}^z + b (S_n^z)^2 \right]
    \label{eq:H}
\end{equation}
with the nearest-neighbor coupling coupling $J>0$, spin
$\mathbf{S}_n^2 = s (s+1)$, and small positive constants $a \ll 1$ and $b \ll 1$, which parameterize the anisotropy. In the large-$s$ limit, neighboring spins are mostly antiparallel, $\thore{\langle}{S}_n\thore{\rangle}\approx - \thore{\langle} \mathbf{S}_{n+1} \thore{\rangle}$ in low-energy states, and long wavelength dynamics of the chain can be understood in terms of the slowly varying unit vector 
$\mathbf{n} = \thore{\langle} \mathbf{S}_{2 n} - \mathbf{S}_{2 n + 1} \thore{\rangle}/s$ in the direction of the local N\'eel order parameter. The dynamics of the field $\mathbf{n}$ follows the nonlinear sigma model with Euclidean action $\mathcal{S} = i \theta Q + \mathcal{S}_0$ (in units of $\hbar$), where $\theta \equiv 2 \pi s$ is referred to as the topological angle. Here,
\begin{equation}
    Q \equiv \frac{1}{4 \pi} \int dx \int d\tau \, \mathbf{n} \cdot (\partial_{x} \mathbf{n} \times \partial_{\tau} \mathbf{n})
    \label{eq:Q}
\end{equation}
is the skyrmion charge of $\mathbf{n}$ that measures how many times $\mathbf{n}(x, \tau)$ wraps the unit sphere as the space and imaginary-time coordinates, $x$ and $\tau$, vary.

QPS are vortex configurations of $\mathbf{n}$ in the two-dimensional Euclidean spacetime. Vortex solutions are characterized by their vorticity $q$ and polarity $p$, which are related to the skyrmion charge as $Q = p q / 2$. When considering a dilute gas of $m$ QPS, the periodic boundary conditions enforce this gas to be vorticity-neutral, $\sum_i q_i=0$.
The resulting topological part of the Euclidean action reduces to~\citeSM{Kim_2016}
\begin{align}
    \mathcal{S}_\theta	&= i \theta \sum_j Q_\thore{j} = i\theta \sum_j p_\thore{j} q_\thore{j} / 2\, , \label{eq:S}
\end{align}
where $j=0,\ldots,m-1$.
For fixed vorticity configuration $\{ q_j \}$, the resulting partition function is summed over the two possible polarities for each QPS, $p_j = \pm 1$, which results in the partition function~\citeSM{Kim_2016}
\begin{align}
        \mathcal{Z} &= \int \mathcal{D} \mathbf{n}(x, \tau) \delta(\mathbf{n}^2 - 1) \exp(- \mathcal{S}_\theta - \mathcal{S}_0) 
        \\ & \nonumber
         \propto \left[ \prod_{j} \cos \left( \frac{\theta q_j}{2} \right) \right] e^{- \mathcal{S}_0( \{ q_j \} )} \, .
    \label{eq:stability}
\end{align}
The prefactor of the partition function distinguishes integer and half-odd-integer $s$. For integer $s$, the topological angle is zero, $\theta = 0$, and thus the prefactor is $1$. Half-odd-integer $s$, however, yields $\theta = \pi$, and the prefactor vanishes when any of vorticities $\{ q_i \}$ is odd. This implies that the QPS destructively interfere for $q=\pm 1$. Thus, in the half-odd integer case, the helices can only be unwound if one has a double winding with $q=\pm 2$.

In the following, we address
the derivation of the topological protection for ferromagnetic spin helices, $J<0$, which can be done analogously to the antiferromagnetic spin helices~\citeSM{Kim_2016} discussed above. Following Ref.~\citeSM{altland_simons_2010_2}, we consider the Wess-Zumino-Witten (WZW) term for the ferromagnetic spin chain,
\begin{align}
    \mathcal{S}_{\rm WZW}(\mathbf{n},\partial_\tau \mathbf{n})&= {i} s\int d\tau \, (1-\cos(\psi))\dot{\phi}.
\end{align}
Here, $(\phi,\psi)$ are two angles parametrizing the unit vector $\mathbf{n}$. This expression is equivalent to~\citeSM{altland_simons_2010_2}
\begin{align}
    \mathcal{S}_{\rm WZW}(\mathbf{n},\partial_\tau \mathbf{n})&=  {i} \frac{C}{4 \pi} \int dx \int d\tau \, \mathbf{n} \cdot (\partial_{x} \mathbf{n} \times \partial_{\tau} \mathbf{n}),
    \label{eq:WZW}
\end{align}
where the coupling constant $C$ obeys the quantization condition $C=4\pi s=2\pi k$, $k\in\mathbb{Z}$. Thus, the $\theta$-term in the antiferromagnetic case~\eqref{eq:Q} is a descendant of the WZW-term in the ferromagnetic case~\eqref{eq:WZW}~\citeSM{altland_simons_2010_2}, and the topological angle $\theta=2\pi s$ can be identified with the coupling constant $C=4\pi s$~\citeSM{altland_simons_2010_2}. This implies that both the antiferromagnetic and ferromagnetic spin chains can be mapped to the same nonlinear sigma model, with the same definitions of the skyrmion charge and vortex configurations as in the previous paragraph. Thus, to derive the topological protection for the ferromagnetic spin helices, we can follow the same derivation as for the antiferromagnetic case~\citeSM{Kim_2016}, arriving at the partition function
\begin{align}
    \begin{split}
        \mathcal{Z} \propto \left[ \prod_{j} \cos \left( \frac{\theta q_j}{2} \right) \right] e^{- \mathcal{S}_0( \{ q_\thore{j} \} )} \,. 
    \end{split}
    \label{eq:Z_F}
\end{align}
Here, $q_i$ is the vorticity of the QPS, just as in the antiferromagnetic case. For half-odd-integer $s$, the prefactor vanishes when any of vorticities $\{ q_i \}$ is odd, which implies that the QPS destructively interfere for $q=\pm 1$. Thus, in the half-odd integer case, the helices can only be unwinded if one has a double winding with $q=\pm 2$.

\section{Time-dependent perturbation theory}
In this section, we conduct the time evolution and averaging over the random fluctuations in the limit of short time steps $\delta t \cdot J /\hbar \ll 1$ and magnetic fluctuations small compared to $1/{\delta t}$, i.e., $h_{\text{max}} \cdot \delta t /\hbar \ll 1$, in order to determine the stability of the helicity of an initial state.

Consider
the Hamiltonian $H(t)$ in Eq.~(1) of the main text describing a spin-$\frac{1}{2}$ chain with $N$ sites perturbed by fluctuating local magnetic fields.
\begin{equation}
    H(t) = H_0 + H_\text{rand}(t) = H_\text{chain} + H_\text{end} + H_\text{rand}(t)
\end{equation}
with dipole-field interaction
\begin{equation}
    H_\text{rand}(t) = \sum_{j=1}^N \vec{S}_j \vec{h}_j(t).
\end{equation}
The chain has $2^N$ eigenstates $\ket{n}$ with eigenenergies $E_n$.
In the interaction picture, the interaction term reads
\begin{align}
    H_{\text{rand},I} (t) =& \sum_{j=1}^N \vec{S}_{I,j}(t) \vec{h}_j(t) 
    \nonumber \\
    =& \sum_{j=1}^N e^{\frac{i}{\hbar}H_0 t} \vec{S}_j e^{-\frac{i}{\hbar}H_0 t} \vec{h}_j(t).
\end{align}
The time evolution operator is conveniently expressed by the Dyson series
\begin{align}
\begin{split}
    U_I(t) = 1 &- \frac{i}{\hbar} \int_{0}^t d\tau H_{\text{rand},I}(\tau) 
     \\ & 
    - \frac{1}{\hbar^2} \int_{0}^t d\tau \int_{0}^\tau d\tau^\prime H_{\text{rand},I}(\tau) H_{\text{rand},I}(\tau^\prime)\\ & + \dots.
\end{split}
\end{align}
Consider an eigenstate $\ket{\Psi}$ and a subspace $A$ spanned by eigenstates of $H_0$ such that $\ket{\Psi} \in A$, we compute the probability to find $\ket{\Psi(t)}$ in $A$.
\begin{equation}
\label{eq:expProjector}
    \bra{\Psi(t)} \mathcal{P}_A \ket{\Psi(t)} = \sum_{\{n\}} \abs{\bra{n}\ket{\Psi(t)}}^2
\end{equation}
with $\mathcal{P}_A$ being the projector onto $A$ and $\{n\}$ a set of orthonormal eigenstates of $H_0$ including $\Psi$ and forming a basis of $A$. We pursue to compute the terms in the sum on the right-hand side of Eq.~\eqref{eq:expProjector}.
\onecolumngridWithRule
\begin{align}
    \bra{n}U_I(t)\ket{\Psi} 
    =\delta_{\Psi n} &- \frac{i}{\hbar} \sum_{j=1}^{N} \sum_{\chi=x,y,z} \int_{0}^t d\tau \bra{n} S_j^\chi \ket{\Psi} h_j^\chi(\tau) 
     \nonumber \\
    &- \frac{1}{\hbar^2} \sum_{m=1}^{2^N} \sum_{j,k=1}^{N} \sum_{\chi,\xi=x,y,z} \int_{0}^t d\tau \int_{0}^\tau d\tau^\prime e^{i (\omega_{nm} \tau - \omega_{\Psi m} \tau^\prime)} \bra{n}S_j^\chi\ket{m}\bra{m} S_k^\xi\ket{\Psi} h_j^\chi (\tau) h_k^\xi (\tau^\prime)
     \nonumber \\
    &+ \dots
\end{align}
Introducing the transition frequencies $\omega_{\Psi n} = (E_\Psi - E_n) /\hbar$ and denoting $\abs{\bra{\Psi}\ket{\Psi(t)}}^2 = \abs{\bra{\Psi}U_I(t)\ket{\Psi}}^2$, we obtain 
\begin{align}
\begin{split}
\label{eq:overlap}
\abs{\bra{n}\ket{\Psi(t)}}^2 = \delta_{\Psi n} &+ \delta_{\Psi n} \frac{2}{\hbar} \sum_{j=1}^{N} \sum_{\chi=x,y,z} \operatorname{Im}\bra{\Psi} S_j^\chi \ket{\Psi} \int_{0}^t d\tau h_j^\chi(\tau) \\
    &+ \frac{1}{\hbar^2} \sum_{j,k=1}^{N} \sum_{\chi,\xi=x,y,z} \bra{n} S_j^\chi \ket{\Psi} \bra{\Psi} S_k^\xi \ket{n} \int_{0}^t d\tau  h_j^\chi(\tau)\int_{0}^t d\tau^\prime h_k^\xi(\tau^\prime) \\
    &- \delta_{\Psi n} \frac{2}{\hbar^2} \sum_{m=1}^{2^N} \sum_{j,k=1}^{N} \sum_{\chi,\xi=x,y,z} \int_{0}^t d\tau \int_{0}^\tau d\tau^\prime \operatorname{Re}\left(e^{i \omega_{\Psi m}(\tau-\tau^\prime)} \bra{\Psi} S_j^\chi\ket{m}\bra{m} S_k^\xi\ket{\Psi}\right) h_j^\chi (\tau) h_k^\xi (\tau^\prime) \\
    &+ \dots
\end{split}
\end{align}

\twocolumngridWithRule

\subsection{Random Sampling}

We draw $M$ random samples of the magnetic field and index them by $\alpha=1, \dots, M$.
We assume that the magnetic field is uncorrelated for different sites and time-differences larger than $\delta t$. Further, we assume that the magnetic field is constant over the time interval
$\left[
    u \delta t,
    (u+1) \delta t
\right]$
with $u = 0, \dots, K-1$ a non-negative integer. The samples are drawn from a distribution of the fields at each lattice site 
within the range $-h_{\text{max}} < h^\chi_j < h_{\text{max}}$ for each field component independently.
The distribution is considered to be independent of time.
As a consequence, we have
\begin{align}
    &\lim_{M\to \infty} \frac{1}{M} \sum_{\alpha=1}^M h_{j,\alpha}^\chi (\tau) = 0 \\
    \label{eq:restriction}
    \begin{split}
        &\lim_{M\to \infty} \frac{1}{M} \sum_{\alpha=1}^M h_{j,\alpha}^\chi (\tau) h_{k,\alpha}^\xi (\tau^\prime) \\
        &=
        \begin{cases}
       \frac{1}{3} h_\text{max}^2 \delta_{j,k} \delta_{\chi,\xi}, & \text{for } u\delta t \leq \tau^\prime < (u+1)\delta t\\
       0, & \text{otherwise},
       \end{cases}
    \end{split}
\end{align}
with $u$ the largest integer such that $u\delta t\leq\tau$. 
With these assumptions, the average of Eq.~\eqref{eq:overlap} over the random samples can be evaluated in a straightforward fashion. 
We define the $\Gamma$-matrix as
\begin{equation}
    \Gamma_{nm} = \sum_{j=1}^{N} \sum_{\chi=x,y,z} \abs{\bra{m} S_j^\chi\ket{n}}^2
\end{equation}
and obtain the total loss of fidelity truncating higher orders of $(h_\text{max} \cdot \delta t / \hbar)$ after reducing $t = \delta t K$ where applicable:
\begin{equation}
    E\left(\bra{\Psi(t)} \mathcal{P}_A \ket{\Psi(t)}\right) 
    =
    1 - \frac{h_\text{max}^2  \delta t}{3\hbar^2} t \sum_{m \notin A} \Gamma_{\Psi m} \operatorname{sinc}^2\left(\frac{\omega_{\Psi m}}{\omega_c}\right)
\end{equation}
with the normalized sine cardinal function $\operatorname{sinc}(x)=\sin(\pi x) / (\pi x)$ and the cutoff frequency $\omega_c \equiv 2\pi \delta t^{-1}$. $E(\cdot)$ denotes the averaging over the sample set for $M \rightarrow \infty$. Note that $K = t / \delta t$ since we always assume $t$ to be an integer multiple of $\delta t$. 

Finally, making the rough approximation that the transition frequencies are always either in the $\abs{\omega_{\Psi m}} \ll \omega_c$ or $\abs{\omega_{\Psi m}} \gg \omega_c$ regime and neglecting the intermediate regime $\abs{\omega_{\Psi m}} \approx \omega_c$, we arrive at the final result for the loss of fidelity of the time-evolved eigenstate
\begin{equation}
\label{eq:loss_final}
    E\left(\bra{\Psi(t)} \mathcal{P}_A \ket{\Psi(t)}\right)  
    \approx 1 - \frac{h_\text{max}^2 \delta t}{3\hbar^2} t  \sum_{m \notin A}^{\abs{\omega_{\Psi m}} < \omega_c} \Gamma_{\Psi m}.
\end{equation}
For a fixed time step $\delta t$, the fidelity decreases linear in time $t$. The smaller $\delta t$, the more $\Gamma$-matrix elements in principle contribute. This effect saturates though for 
$\delta t \ll \hbar/J$, assuming that the energy spectrum is bound, which is the case for the considered spin chains. We then reach at 
\begin{align}
\label{eq:loss_very_final}
    E\left(\bra{\Psi(t)} \mathcal{P}_A \ket{\Psi(t)}\right)  
    \approx 1 - \frac{h_\text{max}^2 \delta t}{3\hbar^2} t  \sum_{m \notin A}\Gamma_{\Psi m},
\end{align}
which results in the equation given in the main text considering that both probabilities add up to one.

\subsection{Lindblad Master equation}
To demonstrate that the stroboscopic time evolution and the chosen uniform random distribution of the perturbing magnetic fields has no physical side effects as compared to a continuous evolution or other random distributions of the random noise, we derive the Lindblad master equation in the limit of small time steps $\delta t$. 
We find that, in first order in $\delta t$, the density matrix $\rho$ evolves as \begin{align}
\label{eqn:Lindblad}
\partial_t \rho(t) = i 
	\left[
		\rho, H 
	\right]
	+ \kappa \sum_{\lambda,j}  
		\left(
			S_j^\lambda\rho S_j^\lambda - \frac{1}{2} 
			\left\{
				{S_j^\lambda}^2,\rho 
			\right\}
		\right), 
\end{align}
with $\kappa = \frac{1}{3} h_{\text{max}} \cdot  {\delta t}$.
The Lindblad master equation generally allows us to directly access the average quantities of the evolved the system. However, relying on the density matrix instead of the state vectors increases the dimension of the implemented matrices by a power of two. Therefore, the~\refEq{eqn:Lindblad} is generally unusable for conducting calculations for as long spin chains as discussed in the main text. For smaller chain lengths, we checked the above Lindblad evolution against the exact numerical implementation of the Schr{\"o}dinger equation discussed in the main text and found that both methods agree within the small stochastic error of the exact numerical simulation.

\bibliographystyleSM{apsrev4-1}
\bibliographySM{Papers,library}

\end{document}